\documentclass[showpacs,preprintnumbers,amsmath,amssymb]{revtex4}
\allowdisplaybreaks
\usepackage{graphicx}
\input epsf
\usepackage{dcolumn}
\usepackage{hyperref}
\hypersetup{colorlinks=true, citecolor=blue, linkcolor=red}
\newcommand{\Ai}{\text{\rm Ai}}
\newcommand{\beq}{\begin{equation}}
\newcommand{\beqn}{\begin{equation*}}
\newcommand{\enq}{\end{equation}}
\newcommand{\enqn}{\end{equation*}}
\newcommand{\eb}{{\rm e}}

\newcommand{\R}{{\mathbb R}}
\newcommand{\N}{{\mathbb N}}

\begin{document}
\allowdisplaybreaks
\title{The Cauchy problem for the generalized hyperbolic Novikov-Veselov equation}
\author{V.A. Yurov}%
\email{vayt37@gmail.com}
\author{A.V. Yurov}%
\email{artyom\_yurov@mail.ru}
\affiliation{Baltic Federal University of I. Kant, Theoretical Physics Department,
 Al.Nevsky St. 14, Kaliningrad 236041, Russia}

\date{\today}
\begin{abstract}
We begin by introducing a new procedure for construction of the exact solutions to Cauchy problem of the real-valued (hyperbolic) Novikov-Veselov equation. The procedure shown therein utilizes the well-known Airy function $\Ai(\xi)$ which in turn serves as a solution to the ordinary differential equation $\frac{d^2 z}{d \xi^2} = \xi z$. In the second part of the article we show that the aforementioned procedure can also work for the $n$-th order generalizations of the Novikov-Veselov equation, provided that one replaces the Airy function with the appropriate solution of the ordinary differential equation $\frac{d^{n-1} z}{d \xi^{n-1}} = \xi z$.

\end{abstract}


\maketitle

\section{Introduction} \label{sec:Intro}
\allowdisplaybreaks
The Novikov-Veselov equation (NV) first came into being almost 30 years ago, in 1984, when Sergey Novikov and Aleksander Veselov have introduced it as a two-dimensional version of the Korteweg-de Vries equation \cite{NV}. Since then, a huge and ever-growing body of works related to the study of NV equations has been established (see, for example, \cite{GN}, \cite{LMSS}, \cite{P} and also \cite{CMMPSS} for a rather extensive review of a recent literature on the subject). One of the most prominent aspects of this interest was an almost universal adoption of an inverse scattering method as a primary tool for conducting the research and finding the exact solutions of NV. However, in this article we wish to discuss an alternative method of solving the Cauchy problem for NV. This method, albeit simple in principle, appears to be deep enough to be applicable to a rather broad class of equations, NV and the 2-dimensional nonlinear Schr\"odinger equation being its two prominent members.


\section{The Moutard Transformation} \label{sec:Moutard}
Let us start by introducing the hyperbolic NV equation:
\beq \label{NV}
\begin{split}
u_t & = u_{xxx}+u_{yyy}+3 \Big((a u)_x + (b u)_y\Big)\\
u_x & = a_y, \quad u_y = b_x.
\end{split}
\enq

This system allows for a Lax pair of the following type:
\beq \label{LA}
\begin{split}
&\Psi_{xy} + u \Psi = 0\\
&\Psi_t = \Psi_{xxx} + \Psi_{yyy} + 3\big(a\Psi_x + b \Psi_y\big).
\end{split}
\enq

If one knows two linearly independent solutions $\Psi_1(x,y,t)$ and $\Psi_2(x,y,t)$ for \eqref{LA}, then one can utilize the famous Moutard transformation to construct a new function $\Psi[1](x,y,t)$ that will serve as a solution to the same equation \eqref{LA} albeit with a new potential $u[1](x,y,t)$. The new potential will then satisfy the relation
\beq \label{u1}
u[1] = u + 2 \partial_x \partial_y \ln \Psi_1.
\enq

Let us assume that $u=a=b=0$. Then the entire system \eqref{LA} simplifies to
\begin{align}
&\Psi_{xy} = 0 \label{LA1} \\
&\Psi_t = \Psi_{xxx} + \Psi_{yyy}. \label{LA2}
\end{align}
The equation \eqref{LA1} can be resolved by separating the variables. The resulting solution will be of a form:
\beq \label{psi1}
\Psi_1(x,y,t)=A(x,t)+B(y,t),
\enq
where $A,B$ are two arbitrary functions that are continuously differentiable by $x$ and $y$, correspondingly. Substituting \eqref{psi1} into \eqref{u1} yields a following post-Moutard form of function $u[1](x,y,t)$:
\beq \label{u}
u[1]=-2\frac{ \partial_x A \cdot \partial_y B}{(A+B)^2}.
\enq
As follows from \eqref{u}, our next goal should lie in ascertaining the exact forms of the functions $A(x,t)$ and $B(y,t)$. This task can be accomplished by looking at the equation \eqref{LA2} which we have ignored so far. We will rewrite it as a standard Cauchy problem by introducing the initial conditions for $A(t,x), B(t,y)$
\beq \label{init}
A(0,x) = \phi(x), \qquad B(0,y) = \Phi(y).
\enq
and rewriting the \eqref{LA2} as a system
\begin{equation} \label{AB}
\begin{split}
A_t & = A_{xxx} + T(t)\\
B_t & = B_{yyy} - T(t),
\end{split}
\end{equation}
where $T = T(t)$ is an arbitrary time-dependent function. The apparently symmetric nature of \eqref{AB} allows us to restrict our attention on just one of the equations therein, namely -- the first one.

We begin by introducing the Fourier transform $\tilde A(p,t)$ of the function $A(x,t)$:
\begin{equation*}
\tilde A(p,t)=\frac{1}{\sqrt{2 \pi}} \int\limits^\infty_{-\infty} A(x,t) \eb^{-i p x} dx.
\end{equation*}

This transformation is handy because of the identity
\beq \label{Afourier}
A(p,t)=\frac{1}{\sqrt{2 \pi}} \int\limits^\infty_{-\infty} \tilde A(p,t) \eb^{i p x} dp,
\enq
which, after being substituted into \eqref{AB}, yields the equation
\begin{equation} \label{ApI}
\int\limits^\infty_{-\infty} \left(\frac{\partial \tilde A}{\partial t} + i p^3 \tilde A - T \right) \eb^{i p x} dp = 0.
\end{equation}
The equation \eqref{ApI} must be satisfied for all $x$ and $p$, and therefore leads to:
\begin{equation} \label{Ap}
\frac{\partial \tilde A}{\partial t} + i p^3 \tilde A = T(t).
\end{equation}

\eqref{Ap} is a nonhomogeneous linear O.D.E. of first order. Its general solution is
\beq \label{Apsol}
\tilde A(p,t)=C(p) \eb^{-i p^3 t} +\int\limits_0^t T(\tau)  \eb^{-i p^3(t-\tau)} d\tau,
\enq
where $C(p)$ is a function, determinable from the initial conditions \eqref{init}. Using the inverse Fourier transform \eqref{Afourier} we come to the following conclusion:
\beq \label{AwithC}
A(x,t)=\frac{1}{\sqrt{2\pi}} \int\limits_{-\infty}^\infty{\left(\int\limits_0^t T(\tau) \eb^{i p^3 \tau} d\tau +C(p) \right)} \eb^{i p x-i p^3 t} dp.
\enq

According to \eqref{init},
\beqn
\phi(x)=\frac{1}{\sqrt{2 \pi}} \int\limits_{-\infty}^\infty C(p) \eb^{i p x} dp,
\enqn
so the unknown $C(p)$ is an inverse Fourier transform of the initial condition $\phi(x)$, i.e.:
\beqn
C(p)=\frac{1}{\sqrt{2 \pi}} \int\limits_{-\infty}^\infty \phi(x) \eb^{-i p x} dx.
\enqn
and we subsequently end up with the following general formula for the function $A(x,t)$:
\beq \label{A}
A(x,t)= \frac{1}{\sqrt{2 \pi}} \int\limits_0^t T(\tau) d\tau \int\limits_{-\infty}^\infty \eb^{i p x - ip^3 (t-\tau)} dp + \frac{1}{2 \pi} \int\limits_{-\infty}^\infty \phi(\xi) d\xi \int\limits_{-\infty}^\infty \eb^{i p (x - \xi) - i p^3 t} dp.
\enq

The \eqref{A} can be further simplified by pointing out the similarity between the integrals with respect to variable $p$ and the {\em Airy function} $\Ai(\xi)$. The Airy function is a particular solution of the eponymous Airy equation:
\beq \label{AiryEq}
\frac{d^2 z}{d \xi^2} = \xi z,
\enq
that has a following integral representation:
\beq \label{Airy}
\Ai(\xi) = \frac{1}{\sqrt{2\pi}} \int\limits_{-\infty}^\infty \eb^{i\left(\frac{t^3}{3}+\xi t\right)} dt.
\enq

Using this fact together with the apparent identity:
\beqn
\frac{1}{\sqrt{2\pi}} \int\limits_{-\infty}^\infty \eb^{-i p a - i p^3 b} dp = \frac{1}{\sqrt[3]{3b}} ~\Ai\left(\frac{a}{\sqrt[3]{3b}}\right),
\enqn
together with the equation \eqref{A} and the similar one written for $B(y,t)$ finally yields:
\beq
\begin{split} \label{ABsol}
A(x,t) &= \int\limits_0^t \frac{T(\tau)}{\sqrt[3]{3(\tau-t)}} ~\Ai\left(\frac{x}{\sqrt[3]{3(\tau-t)}}\right) d\tau + \frac{1}{\sqrt{2\pi}\sqrt[3]{3t}}\int\limits_{-\infty}^\infty \phi(\xi) \Ai \left(\frac{\xi-x}{\sqrt[3]{3t}}\right) d\xi \\
B(y,t) &= - \int\limits_0^t \frac{T(\tau)}{\sqrt[3]{3(\tau-t)}} ~\Ai\left(\frac{y}{\sqrt[3]{3(\tau-t)}}\right) d\tau + \frac{1}{\sqrt{2\pi}\sqrt[3]{3t}}\int\limits_{-\infty}^\infty \Phi(\eta) \Ai \left(\frac{\eta-y}{\sqrt[3]{3t}}\right) d\eta.
\end{split}
\enq

So, we end up with both the solution $\Psi_1=A+B$ of the Lax pair \eqref{LA1}, \eqref{LA2}, and, as a courtesy of Moutard transform \eqref{u1}, with a solution $u[1]$ of the NV equation \eqref{NV} as well. In other words, to find a non-zero solution of the NV equation, it will suffice to start with $u \equiv 0$, impose the boundary conditions \eqref{init} on the Lax pair \eqref{LA1}, \eqref{LA2}, use \eqref{ABsol} to find its solution and conclude the calculations by finding a function $u[1]$ via the Moutard transformation \eqref{u1}. As straightforward as it is, there is one question we should ask: what would happen should we try to {\em invert} the process and instead start out with he boundary conditions for the NV equation itself?

\section{The Cauchy problem for the Novikov-Veselov equation} \label{Sec:Cauchy}

In the previous chapter we have shown that there shall exist a solution $u[1](x,y,t)$ to the NV equation, whose exact form can be derived via the Moutard transformation \eqref{u} from the solutions of the system (\ref{LA1}, \ref{LA2}), provided we are given the initial conditions \eqref{init}. But what would happen if the exact forms of the functions $\phi(x)$ and $\Phi(y)$ are {\em unknown} and we are instead given the initial conditions for the NV equation itself, and would it still be possible to find the required $u[1]$? In other words, is it possible to find an analytic solution to the Cauchy problem for the NV equation provided we only know that the solution has a general structure \eqref{u}? In short, the answer is ``yes''.

Let us start by introducing the set of conditions and boundary conditions for the NV equation:
\beq \label{initNV}
\begin{split}
u[1](x,y,0) &= u_0(x,y)\\
u_0(x,0) &= A_1(x) \\
u_0(0,y) &= B_1(y) \\
A_1(0) &= B_1(0) = C,
\end{split}
\enq
where $C \in \R$ is some constant that is given to us together with the boundary conditions $A_1$ and $B_1$. Since we know that $u[1]$ satisfies the Moutard transformation, we also know that:
\beq \label{u0}
u_0(x,y) = -2 \frac{\phi'(x) \cdot \dot\Phi(y)}{(\phi(x)+\Phi(y))^2},
\enq
where $\phi$ and $\Phi$ are defined as in Sec \ref{sec:Moutard}, and $'$ and $\cdot$ denote the partial derivatives with respect to $x$ and $y$ variables correspondingly. From \eqref{u0} and \eqref{initNV} it immediately follows that
\beq \label{A1C}
\begin{split}
A_1(x) &= -2 \frac{\phi'(x) \cdot \dot \Phi(0)}{(\phi(x)+\Phi(0))^2}\\
B_1(y) &= -2 \frac{\phi'(0) \cdot \dot \Phi(y)}{(\phi(0)+\Phi(y))^2} \\
C &= -2 \frac{\phi'(0) \cdot \dot \Phi(0)}{(\phi(0)+\Phi(0))^2}.
\end{split}
\enq
The first two differential equations in \eqref{A1C} can be easily integrated; for example, the first one after the integration with respect to the variable $x$ yields
\beqn
\frac{\dot \Phi(0)}{\phi(x)+\Phi(0)} = \frac{1}{2}\int\limits_0^x A_1(\xi) d\xi + -\frac{\dot \Phi(0)}{\phi(0)+\Phi(0)},
\enqn
which leads us to the following conclusion:
\beq \label{phi}
\begin{split}
\phi(x) &= \displaystyle{\frac{2 \dot\Phi_0}{\int\limits_0^x{A_1(\xi)d\xi} + \frac{2 \dot\Phi_0}{\phi_0+\Phi_0}}}-\Phi_0\\
\phi'(x) &= \displaystyle{\frac{-2 \dot\Phi_0 A_1(x)}{\left(\int\limits_0^x{A_1(\xi)d\xi} + \frac{2 \dot\Phi_0}{\phi_0+\Phi_0}\right)^2}},
\end{split}
\enq
where we have introduced the notation: $\phi(0)=\phi_0$, $\Phi(0)=\Phi_0$, $\phi'(0)=\phi'_0$ and $\dot\Phi(0)=\dot\Phi_0$. In a similar fashion, the boundary condition $\Phi(y)$ and its derivative will satisfy:
\beq \label{Phi}
\begin{split}
\Phi(y) &= \displaystyle{\frac{2 \phi'_0}{\int\limits_0^y{B_1(\zeta)d\zeta} + \frac{2 \phi'_0}{\phi_0+\Phi_0}}}-\phi_0\\
\dot\Phi(y) &= \displaystyle{\frac{-2 \phi'_0 B_1(y)}{\left(\int\limits_0^y{B_1(\zeta)d\zeta} + \frac{2 \phi'_0}{\phi_0+\Phi_0}\right)^2}}.
\end{split}
\enq
The system (\ref{phi}, \ref{Phi}) depends on four constants: $\phi_0$, $\Phi_0$, $\phi'_0$ and $\dot\Phi_0$. Three of them can be chosen arbitrarily, whereas the fourth one would have to satisfy the equation \eqref{A1C}, namely:
\beqn
-2 \frac{\phi'_0 \cdot \dot \Phi_0}{(\phi_0+\Phi_0)^2} = C.
\enqn
Curiously, this choice does not affect the Cauchy problem of the NV equation in the least, for it can be shown by direct substitution into \eqref{u0} that:
\beq \label{u0last}
u_0(x,y)=\frac{4 C A_1(x) B_1(y)}{\left(\int\limits_0^x \int\limits_0^y A_1(\xi) B_1(\zeta) d\zeta d\xi + 2 C\right)^2},
\enq
i.e. the {\em initial} condition $u_0(x,y)$ depends only on the known {\em initial boundary} conditions $A_1(x)$, $B_1(y)$ and $C$.

We are now ready to answer the question posed in the beginning of this section: provided we know the initial conditions \eqref{u0last}, how do we solve the corresponding Cauchy problem of the hyperbolic real-valued Novikov-Veselov equation? The answer lies in {\em repeating} the Moutard transformation process we described in Sec. \ref{sec:Moutard}! Indeed, since the unknown functions $\phi(x)$ and $\Phi(y)$ satisfy the relations \eqref{phi} and \eqref{Phi}, all we really have to do is substitute them into the system \eqref{ABsol}, derive $A(x,t)$ and $B(y,t)$, and substitute them in equation \eqref{u} to find out the sought after $u[1](x,y,t)$, which will conclude the problem.

Lets summarize everything we have said so far. In order to find an exact solution $u(x,y,t)$ to the hyperbolic real-valued Novikov-Veselov equation
\beq \label{NV1}
u_t = u_{xxx}+u_{yyy},
\enq
with the given initial boundary conditions
\beqn
u(x,0,0) = A_1(x), \qquad u(0,y,0) = B_1(y), \qquad u(0,0,0) = C,
\enqn
that correspond to the initial condition
\beqn
u_0(x,y)=\frac{4 C A_1(x) B_1(y)}{\left(\int\limits_0^x \int\limits_0^y A_1(\xi) B_1(\zeta) d\zeta d\xi + 2 C\right)^2},
\enqn
one shall:
\begin{enumerate}

\item Choose a differentiable function $T(t)$ and four numbers $\alpha$, $\beta$, $\gamma$ and $\delta$ that satisfy the condition,
\beqn
-2 \frac{\gamma \cdot \delta}{(\alpha+\beta)^2} = C.
\enqn

\item Find two support function $\phi(x)$ and $\Phi(y)$ via the formulas
\beqn
\begin{split}
\phi(x) &= \displaystyle{\frac{2 \delta}{\int\limits_0^x{A_1(\xi)d\xi} + \frac{2 \delta}{\alpha+\beta}}}-\beta\\
\Phi(y) &= \displaystyle{\frac{2 \gamma}{\int\limits_0^y{B_1(\zeta)d\zeta} + \frac{2 \gamma}{\alpha+\beta}}}-\alpha.
\end{split}
\enqn

\item Substitute $\phi(x)$ and $\Phi(y)$ into the equations
\beq \label{AB1}
\begin{split}
A(x,t) &= \int\limits_0^t \frac{T(\tau)}{\sqrt[3]{3(\tau-t)}} ~\Ai\left(\frac{x}{\sqrt[3]{3(\tau-t)}}\right) d\tau + \frac{1}{\sqrt{2\pi}\sqrt[3]{3t}}\int\limits_{-\infty}^\infty \phi(\xi) \Ai \left(\frac{\xi-x}{\sqrt[3]{3t}}\right) d\xi \\
B(y,t) &= - \int\limits_0^t \frac{T(\tau)}{\sqrt[3]{3(\tau-t)}} ~\Ai\left(\frac{y}{\sqrt[3]{3(\tau-t)}}\right) d\tau + \frac{1}{\sqrt{2\pi}\sqrt[3]{3t}}\int\limits_{-\infty}^\infty \Phi(\eta) \Ai \left(\frac{\eta-y}{\sqrt[3]{3t}}\right) d\eta.
\end{split}
\enq

\item Substitute the new functions $A(x,t)$ and $B(y,t)$ into the equation
\beq \label{uu}
u = -2\frac{ \partial_x A \cdot \partial_y B}{(A+B)^2}.
\enq
\end{enumerate}

The resulting function $u(x,y,t)$ will be a proper solution of the Cauchy problem since by construction it will satisfy both the NV equation \eqref{NV1}, and the initial conditions $u(x,y,0)=u_0(x,y)$. We would like to emphasize here that this procedure does not involve anything more complicated than partial differentiation and integration and can therefore be used for both the analytic study of the properties of the solutions of NV equation and the corresponding numerical calculations.

\section{Generalization of the method: the higher order equations}

Let us now say a few words about the more general problem. Suppose we have the following operator-type Lax pair:
\beq \label{LAH}
\begin{split}
&\partial_x \partial_y \Psi + u \Psi = 0\\
&\partial_t \Psi = \partial_x^n \Psi + \partial_y^n \Psi,
\end{split}
\enq
where $n \in \N^{+}$ in some non-zero natural number. This system will correspond to a family of Lax equations, with the special case $n=3$ corresponding to the hyperbolic NV equation. It will still allow for the Moutard transformation, and therefore the crux of our discussion would still be applicable for arbitrary $n$. However, one thing that {\em must} change is the exact form of the equations for $A(x,t)$ and $B(y,t)$. The formula \eqref{AB1} is no longer applicable for the general case and should be properly replaced. In order to find out the suitable replacement, we shall separately consider two alternative cases: when $n$ is odd and when $n$ is even.
\newline

{\bf Case 1: Odd $n$.} Let $n=2m+1$, where $m \ge 0$. Following our previous discussion, let us consider the special case $u \equiv 0$. Then the system \eqref{LAH} turns into
\beqn
\begin{split}
&\partial_x \partial_y \Psi = 0\\
&\partial_t \Psi = \partial_x^n \Psi + \partial_y^n \Psi.
\end{split}
\enqn
Since the first equation requires that $\Psi=A(x,t)+B(y,t)$, the system subsequently splits into the following equations:
\beqn
\partial_t A_{2m+1} =\partial_x^{2m+1} A_{2m+1}, \qquad \partial_t B_{2m+1} = \partial_y^{2m+1} B_{2m+1},
\enqn
where for simplicity we have omitted the arbitrary function $T(t)$. Using the Fourier transformation
\beqn
\tilde A_{2m+1}(p,t)=\frac{1}{\sqrt{2 \pi}} \int\limits^\infty_{-\infty} A_{2m+1}(x,t) \eb^{-i p x} dx,
\enqn
we end up with the differential equation
\beqn
\partial_t \tilde A_{2m+1} = (i p)^{2m+1} \tilde A_{2m+1} = i (-1)^m p^{2m+1} \tilde A_{2m+1}.
\enqn
Solving it and returning back to $A(x,t)$ as described in Sec.\ref{sec:Moutard} yields
\beq \label{Aodd}
A_{2m+1}(x,t) = \frac{1}{\sqrt{2\pi}} \int\limits_{-\infty}^{\infty} d\xi ~\phi(\xi) ~\frac{1}{\sqrt{2\pi}} \int\limits_{-\infty}^{\infty} dp ~\eb^{i\left(p(x-\xi)+(-1)^m p^{2m+1} t\right)}.
\enq

As we know, in the special case $m=1$ (i.e. $n=3$) the inner integral in \eqref{Aodd} can be rewritten in terms of the Airy function
\beqn
\Ai(\xi) = \frac{1}{\sqrt{2\pi}} \int\limits_{-\infty}^\infty \eb^{i\left(\frac{t^3}{3}+\xi t\right)} dt = \sqrt{\frac{2}{\pi}} \int\limits_0^\infty \cos{\left(\xi t +\frac{t^3}{3}\right)} dt,
\enqn
which serves as a solution to the Airy equation
\beqn
\frac{d^2 z}{d \xi^2} = \xi z,
\enqn
and is easily derived using either Fourier or Laplace transformation \footnote{In case of the Laplace transformation the contour of integration must be chosen lying inside of a sector where $\cos(3\theta) >0$}.

Similarly, it is easy to show that one of a solutions to a more general equation
\beqn
\frac{d^{2m} z}{d \xi^{2m}} = \xi z,
\enqn
will be a {\em higher-order generalization} of the Airy function \footnote{We would like to remind the reader that in literature the term {\em generalized Airy function} is commonly assigned to the solutions of the second order O.D.E. $w''(x)=x^n w(x)$; hence the addition of the term {\em higher-order} in our case is necessary to avoid a possible confusion.}:
\beq \label{Airyodd}
\Ai_{2m+1}(\xi) = \frac{1}{\sqrt{2\pi}} \int\limits_{-\infty}^\infty \eb^{i\left(\xi t - \frac{(-1)^m}{2m+1} ~t^{2m+1}\right)} dt = \sqrt{\frac{2}{\pi}} \int\limits_0^\infty \cos{\left(\xi t - \frac{(-1)^m}{2m+1} ~t^{2m+1}\right)} dt,
\enq
which means that the required functions $A_{2m+1}$ and $B_{2m+1}$ can be derived from the initial conditions $\phi(x)$ and $\Phi(y)$ by the following formulas:
\beq \label{ABodd}
\begin{split}
A_{2m+1}(x,t) &= \frac{1}{\sqrt[2m+1]{(2m+1)t}} \int\limits_{-\infty}^\infty d\xi ~\phi(\xi) ~\Ai_{2m+1}\left(\frac{\xi-x}{\sqrt[2m+1]{(2m+1)t}}\right),\\
B_{2m+1}(x,t) &= \frac{1}{\sqrt[2m+1]{(2m+1)t}} \int\limits_{-\infty}^\infty d\zeta ~\Phi(\zeta) ~\Ai_{2m+1}\left(\frac{\zeta-y}{\sqrt[2m+1]{(2m+1)t}}\right).
\end{split}
\enq
\newline

{\bf Case 2: Even $n$.} Let $n=2m$, where $m \ge 0$. This time let us utilize not a Fourier but a Laplace transform:
\beqn
A_{2m}(x,t)=\int\limits^\infty_{-\infty} \tilde A_{2m}(p,t) \eb^{p x} dx.
\enqn
the equation for $\tilde A(p,t)$ is
\beqn
\frac{\partial \tilde A_{2m}}{\partial t} = p^{2m} \tilde A_{2m},
\enqn
so the required function $A(x,t)$ will satisfy the equation
\beq \label{Aodd}
A_{2m}(x,t) = \int\limits_{-\infty}^{\infty} d\xi ~\phi(\xi) ~\int\limits_{-\infty}^{\infty} dp ~\eb^{p(x-\xi)+p^{2m} t}.
\enq

It is not difficult to show that the Laplace transformation method applied to the ordinary differential equation
\beqn
\frac{d^{2m-1} z}{d \xi^{2m-1}} = \xi z,
\enqn
will yield a following solution
\beq \label{Airyeven}
\Ai_{2m}(\xi) = \int\limits_{-\infty}^\infty \exp \left(\xi t - \frac{t^{2m}}{2m}\right) dt,
\enq
and so the even case produces the formulas that are quite similar to the old ones, namely:
\beq \label{ABeven}
\begin{split}
A_{2m}(x,t) &= \frac{1}{\sqrt[2m]{-2 m t}} \int\limits_{-\infty}^\infty d\xi ~\phi(\xi) ~ \Ai_{2m}\left(\frac{x-\xi}{\sqrt[2m]{-2mt}}\right),\\
B_{2m}(x,t) &= \frac{1}{\sqrt[2m]{-2mt}} \int\limits_{-\infty}^\infty d\zeta ~\Phi(\zeta) ~\Ai_{2m}\left(\frac{y-\zeta}{\sqrt[2m]{-2mt}}\right).
\end{split}
\enq
Note the appearance of a negative sign under the root in \eqref{ABeven}, which serves as a indication of an ill-posedness of our problem for $t>0$.

\end{document}